\documentstyle[pre,aps,multicol,epsf]{revtex}

\begin{document}

\title{Generic Model of Morphological Changes 
in Growing Colonies of Fungi}

\author{Juan M. L\'opez$^{1,}$\cite{mail1}
and Henrik J. Jensen$^2$\cite{mail2}}

\address{$^1$ Instituto de F{\'\i}sica de Cantabria,
Consejo Superior de Investigaciones 
Cient{\'\i}ficas--Universidad de Cantabria, 
E-39005 Santander, Spain}

\address{$^2$ Department of Mathematics, 
Imperial College of Science, Technology and Medicine, 
180 Queen's Gate, London SW7 2BZ, United Kingdom.}

\maketitle

\begin{abstract}
Fungal colonies are able to exhibit different morphologies 
depending on the enviromental conditions. This allows 
them to cope with and adapt to external changes. 
When grown in solid or semi-solid media the bulk of the 
colony is compact and several morphological transitions 
have been reported to occur as the external conditions 
are varied. Here we show how a unified simple mathematical
model, which includes the effect of the accumulation of 
toxic metabolites, can account for the morphological changes 
observed. Our numerical 
results are in excellent agreement with experiments carried out 
with the fungus {\em Aspergillus oryzae} on solid agar.
\end{abstract}


\begin{multicols}{2}
\narrowtext

\section{Introduction}
Since the pioneering work of 
D'Arcy Thompson \cite{darcy} at the beginning
of this century understanding the origin of complex structures 
involving living organisms has been a challenge for physicists 
and biologists alike. In particular the patterns formed 
in growing colonies of microorganisms, 
like bacteria 
\cite{cooper,fuji,matsushita,jacob94,matsuyama,jacob97} 
and fungi 
\cite{obert,jones,mihail,bolton,donnelly,ritz,matsuura92,matsuura93,sams}
have attracted much interest because
they exhibit properties akin to those 
in inanimate systems far from equilibrium \cite{jacob90}.
Early mathematical models of fungal 
growth \cite{trinci,prosser},
considered fungi as purely additive assemblages of
discrete individual hyphal units that duplicate at
regular intervals and homogeneously.  
These models predict the crossover from  
exponential growth to linear growth 
of the peripheral zone as the colony matures. 
However, they are unable to account for
many of the observable macroscopic properties
(as morphological changes for instance) 
exhibited by fungi.

Fungal growth experiments on agar plates have shown 
\cite{obert,jones,mihail,bolton,donnelly,ritz,matsuura92,matsuura93,sams}
that filamental growth leads to a
fractal colony with patterns that have much in 
common with those observed in diffusion-limited 
aggregation (DLA) processes \cite{witten,sander}. 
Similar branched morphologies have also been
observed in bacterial growth
\cite{cooper,fuji,matsushita,jacob94} 
where the relevance of DLA was nicely demonstrated
\cite{fuji,matsushita}.
Sophisticated communication mechanisms,
including chemotactic signals and feedback response, 
were later proposed \cite{jacob94}
in order to understand the origin of 
branched growth in bacteria. A similar mathematical 
modelling might be
adequate to explain the DLA-like patterns 
of hyphae. However, quantitative measurements of the 
fractal dimension of mycelia 
\cite{ritz,obert} have shown that 
the dimension varies in time and from 
one species to another.
Also, the growth mechanism of DLA clusters seems rather
different from the hyphal extension of fungi.
Some reaction-diffusion models have also been studied
\cite{regalado,davidson}, which seem to capture esential 
features of the filamental fungi. 

In contrast, under appropriate environmental
conditions, colonies of many fungal species are 
also able to display 
a compact phase \cite{gow,carlile}
with uni-cellular non-cooperative growth
and generate non-DLA patterns. 
In this case, transitions between different morphologies 
are also observed \cite{matsuura92,matsuura93,sams} 
when the relevant parameters in the growth 
conditions are varied. A mathematical model able 
to include all these fungal patterns on solid 
media has remained elusive up to now \cite{sams,lopez}.
Here we show that a simple stochastic model 
containing some generic
biological mechanisms is able to reproduce a 
variety of patterns 
observed in the growth of fungal colonies 
on solid agar plates. 
The model takes into account the effect of 
the accumulation of waste products, which play 
an important role as inhibitors for cell division.

Matsuura and Miyazima \cite{matsuura92,matsuura93}
carried out experiments with the fungus 
{\em Aspergillus oryzae} on agar plates of 
different stiffness and with different concentrations 
of nutrients (glucose). 
They observed that the fungal colonies on solid media 
formed a variety of shapes (see Figure 1).
In nutrient-rich conditions, the colonies showed a 
homogeneous and smooth growth front (Fig. 1a). 
In contrast, growth in glucose-poor conditions 
resulted in colonies with rougher surfaces (Fig. 1b). 
The physiological ability of the fungus to
divide and grow is diminished at lower 
temperatures \cite{gow,carlile},
when even morphological instabilities (Fig. 1c)
or an incipient branched growth (Fig. 1d) can be 
observed depending on the nutrient concentration. 
Similar patterns have recently been  
obtained in experiments with colonies of the yeast
{\em Pichia membranaefaciens} on solidified agarose 
\cite{sams}. How can these morphological changes be 
qualitatively described and understood within
one simple mathematical model?.

\section{The Model}
It is known that
metabolic products generated by fungal cells during the 
degradation of nutrients diffuse through the medium 
and may act as impurities inhibiting 
further cell division \cite{gow,carlile,sams}.
We have considered the effect of toxic metabolites 
on the morphology of the colony.
We have studied a model 
in which the growth probability at every site is 
coupled with an inhibitory field generated by the 
fungus itself. 
The model is defined on a two dimensional 
lattice as follows. At time $t$ a lattice 
site can be either {\em occupied} by a fungal 
cell or {\em vacant}. 
Two quantities are assigned to every site 
${\vec x}$ in the lattice: 
(1) the total age of the nearest neighbour 
cells $A({\vec x},t)$ of that site
and (2) the field $c({\vec x},t)$ containing 
the concentration of waste products. 
The spread of inhibitors from a single cell 
situated at $\vec x_0$ is modelled by means of 
a diffusion equation with a source:
\begin{equation}
\label{diff-inh}
\frac{\partial c(\vec x,t)}{\partial t} = D \: \nabla^2 c(\vec x,t)
+ s \: \delta(\vec x - \vec x_0) \: \Theta(t-\tau_0),
\end{equation} 
where $\Theta$ is the step function 
($\Theta(u) = 1$ for $u \ge 0$ and 
$\Theta(u) = 0$ for $u < 0$). 
$D$ is the diffusion constant of the metabolites, 
$s$ is the inhibitor chemicals' production rate, 
and $\tau_0$ is the time at 
which the polluting cell
at $\vec x_0$ was born. The total density of 
waste products at any vacant site at time $t$ is 
calculated by adding the contributions coming from 
every occupied site in colony situated 
within a distance $d$. 

Colony growth occurs because of the division of 
individual cells, thus only nearest neighbours of 
occupied sites have a chance
of becoming occupied. We assume that  
the probability for a vacant site
${\vec x}$ of being occupied in the next time 
step increases with the total age $A({\vec x},t)$ 
of the occupied nearest neighbours of that vacant 
site $\vec x$. Therefore, at every time step, 
we assign a growth probability $P({\vec x},t)$ 
to every vacant site ${\vec x}$ given by
\begin{equation}
\label{grow-prob}
P({\vec x},t) = F [\theta \: A({\vec x},t) \: 
\phi(c)],
\end{equation}
which depends on the two fields 
$A({\vec x},t)$ (total age of 
neighbouring occupied sites) and $c(\vec x,t)$ 
(waste concentration) introduced above. 
In Eq.(\ref{grow-prob}),
$F$ is an arbitrary monotonous increasing function 
satisfying $F(0) = 0$ and $F(\infty) = 1$ 
(properties required for $P$ to be a probability). 
The results do not depend on the detailed form of $F$ 
and in the following we use $F(x) = {\rm tanh}(x)$ for 
simplicity. The suppression of cell 
division due to the accumulation of metabolites is 
represented by $\phi(c)$ that is a decreasing function 
of the concentration of waste chemicals. In order to see 
a significant effect of the inhibitors, the decay of 
$\phi$ with $c$ must be rapid enough. In the simulations 
we are presenting in this paper $\phi(c) = {\rm exp}(-c/c_0)$ 
was assumed, $c_0$ being a concentration threshold above 
which the effect of inhibitors becomes important. 
Other simple functional forms may be used 
and similar patterns are obtained provided that $\phi(c)$ 
decays rapidly enough. Finally, the external parameter 
$\theta$ controls directly the
growth rate and is associated with the 
food supply (media richer 
in nutrients correspond to larger values of $\theta$).

\section{Results}
We carried out extensive simulations of the model. 
The simulations were performed on a two dimensional 
triangular lattice with open boundary conditions,
simultaneously updating Eqs.(\ref{diff-inh}) and 
(\ref{grow-prob}) in every time step. The position of 
the front is determined and
the height $h(x,t)$ at every column $x$ is used to 
study the front dynamics. In order to mimic the experiments, 
the growth is initiated from a straight line of randomly 
occupied sites at $h(x,0) = 0$. The dynamics of the growing 
front is characterized by the height fluctuations, which are 
measured by means of the front width
over the total system of size $L$,
$W(L,t) = \sqrt{\langle (h - \langle 
h \rangle)^2\rangle}$.
The use of other quantities to measure the width is helpful
when instabilities or overhangs appear, like for instance 
the difference $Max[h] - Min[h]$
between the maximum and minimum heights, 
and gave similar results. 

The model has two external parameters,
namely: the growth 
rate $\theta$, and the concentration threshold $c_0$ for 
inhibitory effects. 
These are to be thought of as 
effective parameters and their relation to experimental
variables such as temperature, concentration of 
nutrients, {\it etc.} may not be a simple one. 
It is to be noted that,
if experimental conditions change, for instance 
in such a way that the absorption of inhibitors 
by the fungal cells is enhanced, the effect may 
be well represented by either a lower threshold 
$c_0$ or a wider spread distance $d$
(or even a combination of both).

We performed computer simulations of the model 
for different values of $\theta$ and $c_0$.
In the following we discuss the
different front morphologies that are obtained
depending on the values of these two parameters,
and how they are related to experimental observations.
In Fig. 2 we present a summary of 
the resulting numerical fungal patterns. 

\subsection{High threshold for inhibitory effects}
The first morphological transition occurs
in the limit $c_0 \to \infty$, in which
the inhibitors become irrelevant ($\phi(c) \approx 1$).
We found that the surface of the colony is 
flat (Fig. 2a) when inhibitors have little effect 
(for $c_0 \gg 1$) and the medium is rich in
nutrients. Conversely, the growth front
becomes rougher and Eden-like (Fig. 2b) for a poorer 
medium. In fact, a quantitative measure 
of the roughness (see below)
indicates that the fractal dimension of the 
front is $1.5$ \cite{lopez} 
as in the Eden model \cite{family}. 
These morphological phases are 
separated by a continuous phase transition 
at a critical value of $\theta=\theta_c$. 
For large growth rates (rich medium)
$\theta > \theta_c$ the front is flat, 
$W \sim {\rm const}$. In contrast,
if the growth rate is reduced below $\theta_c$
(poor medium) the front becomes rougher
and exhibits the dynamical scaling behaviour 
typical of a scale-free 
roughening process \cite{baraba,family}.
As can be seen in Fig. 3, in the rough phase
the width grows like a power law 
$W \sim t^{\beta}$ until a stationary
regime is reached and $W \sim L^{\alpha}$, 
where $L$ is the system
size. Our determination of the
time exponent $\beta = 0.24 \pm 0.02$ and the 
roughness exponent $\alpha = 0.46 \pm 0.05$ indicates 
that the model belongs to the Edwards-Wilkinson 
universality class \cite{ew} in this phase. 
The computation of the roughness exponent $\alpha$ 
allows us to determine the fractal dimension $D_f$ 
of the front through the relation $D_f=2-\alpha \simeq 1.54$, 
which is consistent with Eden-like growth.
This morphological transition
from rough to flat growth has recently been studied 
in detail by us \cite{lopez}
for a much simpler version 
of this model, which corresponds to the 
limit $c_0 \to \infty$.
We found that $\theta_c = 0.183 \pm 0.003$
in the limit case of $c_0 = \infty$.
When approaching the roughening
transition from the rough phase ($\theta < \theta_c$) 
to the flat phase ($\theta > \theta_c$), a diverging 
correlation lenght $\xi \sim |\theta - \theta_c|^{-\nu}$ 
appears. The critical exponent was found to be $\nu = 1.10$. 
At the phase transition the scaling
exponents can be calculated by mapping the problem onto 
directed percolation (see Ref. \cite{lopez} for 
further details). 
Experimental realizations of transitions from a flat
to a rough front can been observed in the 
Matsuura and Miyazima experiments shown in Figure 1a-1b.
As our model predicts the richer the medium the flater 
the front.

\subsection{Low threshold for inhibitory effects}
If the effect of toxic
inhibitors on the fungus is enhanced,
further interfacial instabilities in the front appear.
Lower temperatures are less favorable 
for the physiological activity of the fungus
\cite{gow,carlile} 
and its capacity to divide. One can model this effect by
a smaller threshold $c_0$ for the inhibition 
of cell division.
One can observe (Fig. 2c) that,
as $c_0$ is decreased, 
deep grooves and overhangs develop in the colony and
some regions with high concentration of inhibitors
are eventually surrounded by the colony front.
The presence of a unstable growth dynamics can be  
realized by the appearance of an extremely 
rapid (exponential) growth of the interface width 
$W(L,t)$ in this regime, as
shown in Fig. 4. This regime is to be compared with the 
patterns observed in the expriments (Fig. 1c).

Finally, in the worst of the situations, 
the growth becomes very localized and only 
the dominant filaments branch out into the 
empty medium (Fig. 2d). This is due to the building up
of extremely high concentrations of waste products in some
regions that the fungus cannot invade any longer. 
There is qualitative agreement with
the experiment shown in Fig. 1d. 

The four major morphological phases are
summarized in the 
qualitative phase diagram in Fig. 5.
The solid line that separates the rough 
from the flat phase in Fig. 5 
has recently been shown to be
a nonequilibrium continuous phase transition 
and can be mapped into
directed percolation \cite{lopez}. 
However, we were unable to find a sharp 
transition between the remaining phases. 

\section{Discussion}
Our results indicate that the morphologies 
generated during the growth of fungal 
colonies on solid agar
can be modeled by taking into account
(1) the building up of 
a concentration of waste inhibitors in the
medium and (2) the provision of nutrients.
We have studied a simple model which allows us to 
understand the interplay between these two mechanisms.
The morphological tranformations that we have found can
be compared with recent experiments on pattern formation
in fungal growth. 

In our model, a constant growth rate $\theta$ 
has been assumed. The effect of the exhaustion
and diffusion of
nutrients can be modeled by replacing $\theta$ 
by a diffusive field, which is depleted at a certain
rate at sites occupied by fungal cells.
This would have a major impact on the front dynamics
as the front will eventually stop growing
because of agar exhaustion. 

Our model approach to pattern formation in fungal
colony growth is stochastic and includes in a 
natural way the random nature of cell division. 
This together with the
accumulation of inhibiting chemicals produces 
some of the different morphological
phases observed in experiments. This is in contrast to other 
existing models based upon non-linear diffusion, in which evolution
is deterministic and patterns arise from nonlinear instabilities.
They represent a different level of description of the same phenomena.
Both type of models coexist at different levels of description.

Finally, while our model can account for the morphological
changes observed in some fungi growing in agar culture,
we should mention that this model may not be appropiate 
for describing the morphological patterns observed in
other fungal species. Our results compare well with the
fungus {\em Aspergillus oryzae} which, like many members
of this genus (and also the genus {\em Penicillum}),
form dense colonies that expand only very slowly and seem
to be self-inhibitory.

The authors are in debt to S. Matsuura 
for kindly providing us photographs of their experiments. 
We thank T. Sams for useful discussions on his work and
M.J. Carlile, S. Matsuura, and E. Calle
for helpful comments and correspondence. 
This work has been supported by a grant from the 
European Commission (Contract No. HPMF-CT-99-00133)).

\end{multicols}

\begin{figure}
\centerline{
\epsfxsize=7.5cm
\epsfbox{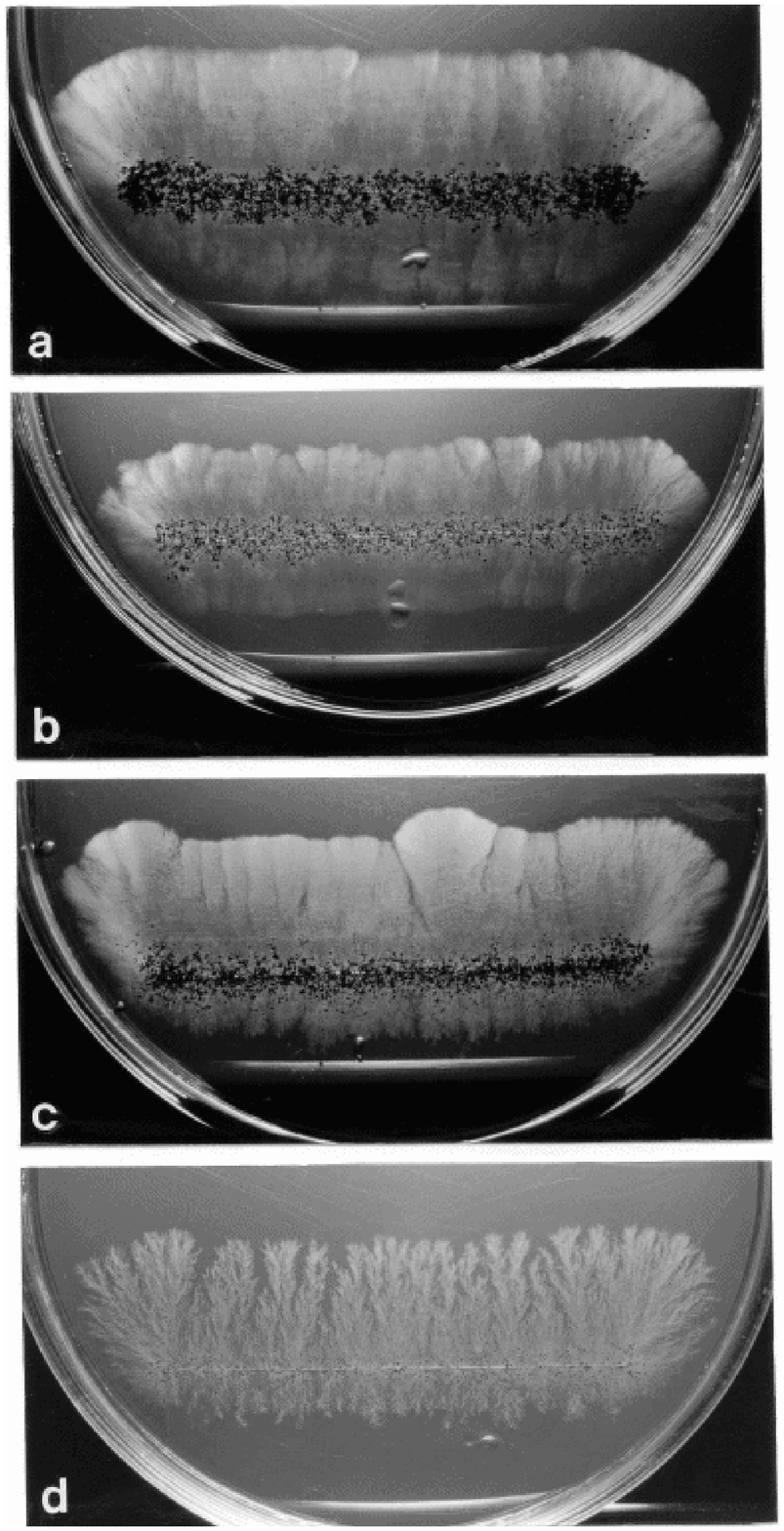}}
\caption{Observed patterns of 
colonies of {\em Aspergillus oryzae} grown on solid medium 
($1.5$ wt\% of a Czapek synthetic agar in $25$ ml of sterile medium). 
Experiments were carried out in media 
with two different concentrations of nutrients: 
$0.1$ wt\% glucose for a nutrient-rich medium and 
$0.01$ wt\% glucose in nutrient-poor conditions. 
(a) Smooth colony $10$ days after inoculation at $24^0$C in a
nutrient-rich medium; 
(b) patterns $8$ days after inoculation at $24^0$C 
in a poor medium, where the colony developed a rough front. 
(c) $30$ days at $18^0$C in a rich medium. Note the tip splitting
dynamics and groovy structure of the colony associated with the
existence of a morphological instability.
(d) Prominent filamental patterns of a 
colony grown in a nutrient-poor medium 
after $35$ days at $18^0$C. Photographs kindly provided by 
S. Matsuura. 
}
\end{figure}

\begin{figure}
\centerline{
\epsfxsize=7.5cm
\epsfbox{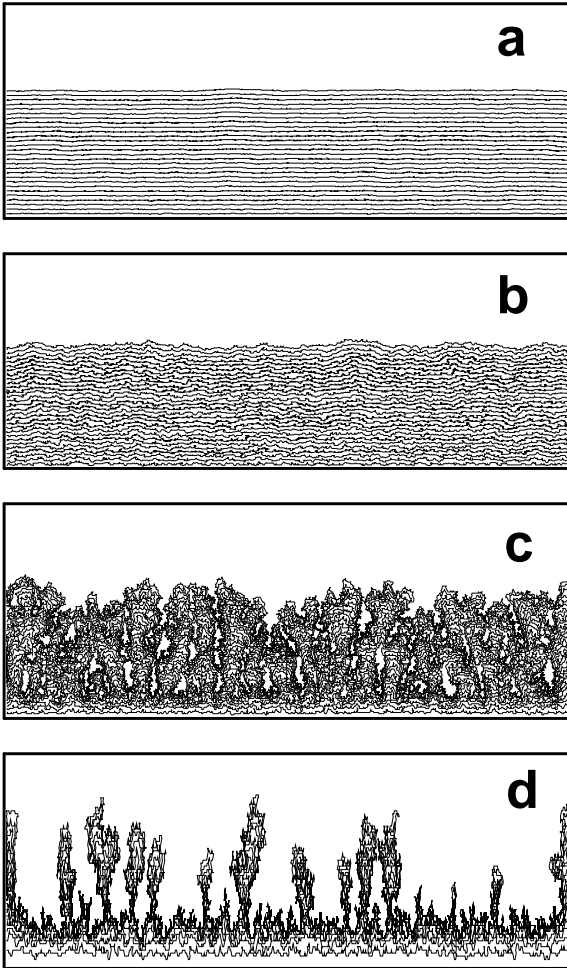}}
\caption{Fungal patterns obtained by numerical simulation 
of the model on a triangular lattice of lateral size $N=500$. 
Each front corresponds to the position of the 
colony surface at equal time intervals. 
When the threshold $c_0$ is large ($c_0 = 5.0$), 
two different morphologies appear: (a) the colony is flat 
in a nutrients-rich medium, $\theta=0.1$ 
and (b) rough in the nutrients-poor case, $\theta=0.01$.
For small values of the threshold $c_0$, 
the colony exhibits a morphological 
instability as in (c) for $c_0=1.0$ 
; or prominent filamental growth as in 
(d) for $c_0=0.7$.
In the simulations shown, the production rate of toxic metabolites
and the diffusion constant are fixed to $s=0.05$ 
and $D=1$ respectively. Sites within a distance $d=16$ 
are considered to calculate the concentration of inhibitors
at any site. 
}
\end{figure}

\begin{figure}
\centerline{
\epsfxsize=7.5cm
\epsfbox{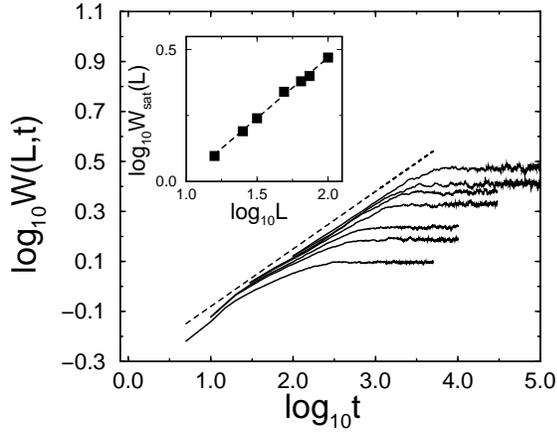}}
\caption{Interface width of the front {\it vs.} time in the rough
phase ($\theta = 0.01$) for different system sizes
$L= 16, 25, 32, 50, 64, 75, 100$ . The slope of the dotted line
corresponds to the time exponent $\beta =0.24 \pm 0.02$.
In the inset the values of the width at saturation
are plotted {\it vs.} system size. The dotted line fits
the data and gives the roughness
exponent $\alpha = 0.46 \pm 0.05$.}
\end{figure}

\begin{figure}
\centerline{
\epsfxsize=7.5cm
\epsfbox{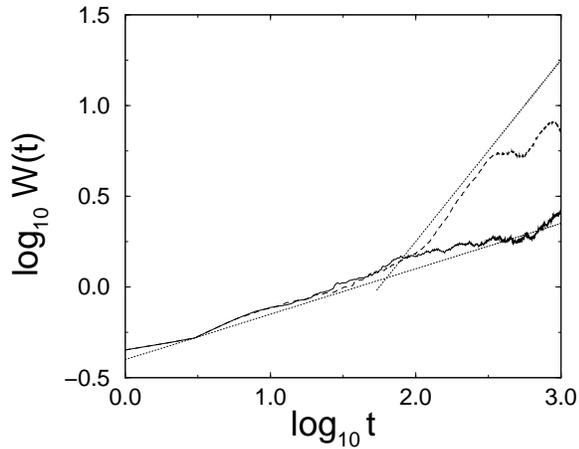}}
\caption{Time behaviour of the interface 
width in the rough (solid line) and unstable phases (dashed line). 
Note the change from power law bahaviour $W(t) \sim t^\beta$ to 
exponential growth $W(t) \sim {\rm exp}\: t$ in the unstable case.
The lines are to guide the eye and have slopes $0.25$ and $1$.
The system size was $L=500$. The external 
parameters are 
$\theta = 0.01$ and $c_0=5$ (rough phase) and 
$\theta = 0.01$ and $c_0=1$ (grooves' phase), corresponding
to Figs. 2b and 2c respectively.}
\end{figure}

\begin{figure}
\centerline{
\epsfxsize=7.5cm
\epsfbox{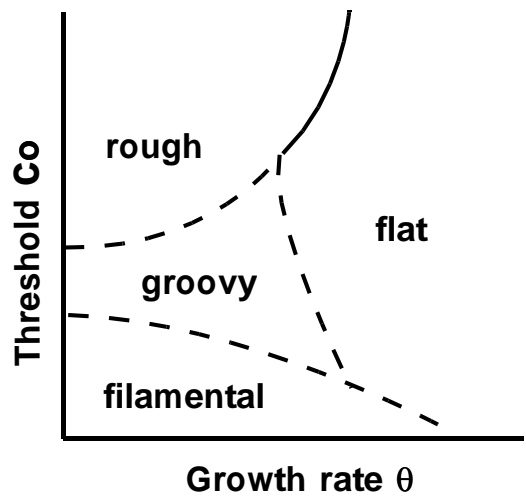}}
\caption{Sketch of the phase diagram showing the different
morphologies observed in the model. The solid line
corresponds to the continuous phase transition studied
in Ref.[26].}
\end{figure}

\end{document}